\documentclass[12pt]{article}
\newcommand{\be}{\begin{equation}}
\newcommand{\ee}{\end{equation}}
\newcommand{\bea}{\begin{eqnarray}}
\newcommand{\eea}{\end{eqnarray}}

\usepackage[icelandic,english]{babel}
\usepackage[T1]{fontenc}
\usepackage{textcomp}
\usepackage{type1cm,amsmath,amssymb,color,graphics,amscd,amsfonts,mathrsfs,epsf,indentfirst}
\usepackage{epsfig}
\usepackage{bm}
\usepackage{subfigure}
\usepackage{multirow,graphicx}
\usepackage{cite}

\newcommand{\tho}{{\textrm{\th}}}
\newcommand{\es}{{\textrm{\dh}}}
\newcommand{\thop}{\tho'}

\makeatletter

\@addtoreset{equation}{section}
\makeatother

\title{
Instability of higher dimensional extreme black holes}

\author{Keiju Murata \\ {\small 
DAMTP, Centre for Mathematical Sciences, Wilberforce Road, Cambridge CB3
0WA, UK}\\
{\small 
Yukawa Institute for Theoretical Physics, Kyoto University,
Kyoto, 606-8502, Japan}
}

\begin{document}
\maketitle

\begin{abstract}
We study linearized gravitational perturbations of extreme black hole
 solutions of the vacuum Einstein equation in any number of
 dimensions. We find that the equations governing such perturbations can
 be decoupled at the future event horizon.
Using these equations, we show that transverse derivatives of certain
 gauge invariant quantities blow up at late time along the horizon if
 the black hole solution satisfies certain conditions.
We find that these conditions are indeed satisfied by many extreme
 Myers-Perry solutions, including all such solutions in five dimensions.
\end{abstract}

\section{Introduction}

Extreme black holes have theoretical importance 
in  
understanding of quantum
theory of gravity. 
For example, Bekenstein-Hawking entropy of supersymmetric black
holes was explained by counting BPS states in the view of string theory~\cite{Strominger:1996sh}.
Furthermore, a duality called the Kerr/CFT correspondence between
extreme black holes and a two-dimensional conformal field theory was
proposed~\cite{Guica:2008mu},
and the entropy of the black holes was reproduced as the
statistical entropy of the dual CFT.

Recently, it was shown that 
extreme Reissner-Nordstr\"{o}m and Kerr black holes 
are classically unstable against
test scalar field 
perturbations~\cite{Aretakis:2011ha,Aretakis:2011hc,Aretakis:2011gz,Aretakis:2012ei}. 
Subsequently, the proof is extended to all other extreme black
holes~\cite{Lucietti:2012sf}.
They showed that the second transverse derivative
blows up at the horizons as $\partial_r^2 \phi\sim v$, 
where $\phi$ is the scalar field and 
we took the ingoing Eddington-Finkelstein coordinates, $(v,r)$.
For extreme Kerr black holes, 
the similar instabilities were also found in 
gravitational and electromagnetic 
perturbations~\cite{Lucietti:2012sf}.

We arise following questions: 
Are all extreme black holes unstable against gravitational or
electromagnetic perturbations? 
If not, what is the condition for
the instability? 
In this paper, we address these questions by 
studying the perturbations of any extreme black holes. 
We use Geroch-Held-Penrose (GHP) formalism in higher dimensions developed
in~Refs.\cite{Durkee:2010xq,Durkee:2010qu} to study the perturbations.
So, in section.\ref{GHP}, 
we give a brief review of the GHP formalism.
We introduce gravitational, electromagnetic and scalar field 
perturbation equations based on the formalism. 
They are regarded as 
higher dimensional analogues of Teukolsky equations 
although they are not decoupled equations 
for gravitational and electromagnetic perturbations in general.
In section.\ref{EXBHinGHP}, we introduce the most general expression for
extreme black hole and express them in the view of the GHP formalism.
In section.\ref{scalar_pert}, 
we study the scalar field perturbation.
Although the scalar field perturbation on any extreme black holes
has been already studied in~Ref.\cite{Lucietti:2012sf}, 
we revisit the problem using the GHP formalism.
We find that 
all extreme black holes are unstable against scalar field perturbations
as shown in~Ref.\cite{Lucietti:2012sf}.
In section.\ref{em_pert}, we study electromagnetic perturbations.
We find that, near the horizon, 
electromagnetic perturbations satisfy decoupled equations. 
Using the decoupled equations, we show that the perturbations do not
decay along the future event horizon if a certain operator on the
horizon has a zero eigenvalue. 
In section.\ref{grav_pert}, 
we study gravitational perturbations.
By the similar way as electromagnetic perturbations,
we can show the non-decay of the gravitational perturbations 
if a horizon operator has a zero eigenvalue. 
In addition to that, 
if the background geometry is algebraically special, 
the first or second  transverse derivatives of the perturbation variables
blow up along the horizon.
The eigenvalues for the horizon operators have been calculated for some
extreme black holes. 
In section.\ref{UnsEX}, we see that there are zero eigenvalues
in the horizon operators for all higher dimensional extreme black holes
with zero cosmological constant as far as we calculated.
The final section is devoted to discussions.

\section{Geroch-Held-Penrose formalism in higher dimensions}
\label{GHP}

We study the perturbation of the general extreme black holes 
using the Geroch-Held-Penrose (GHP) formalism in higher dimensions 
developed in~Refs.\cite{Durkee:2010xq,Durkee:2010qu}.
In this section, we give a brief review of the GHP formalism.
In the formalism,
we use a null basis $\{e_0,e_1,e_i\}=\{\ell,n,m_i\}$ $(i=2,\cdots,d-1)$ 
which satisfies
\begin{equation}
\label{null_orth}
 \ell^2=n^2=\ell\cdot m_i=n\cdot m_i=0\ ,\quad
 \ell\cdot n=1\ ,\quad
 m_i\cdot m_j=\delta_{ij}\ .
\end{equation}
We define the covariant derivatives of basis vectors as
\begin{equation}
\label{LNMdef}
 L_{ab}=\nabla_b \ell_a\ ,\quad
 N_{ab}=\nabla_b n_a\ ,\quad
 M^i_{ab}=\nabla_b m_{ia}\ ,
\end{equation}
and 
\begin{equation}
 \rho_{ij}=L_{ij}\ ,\quad
 \tau_i=L_{i1}\ ,\quad
 \kappa_i=L_{i0}\ .
\end{equation}
The orthogonal relations~(\ref{null_orth})
are invariant under spins, boosts and null rotations defined
as follows. Spins are local $SO(d-2)$ rotations of the spacial basis $\{m_i\}$:
\begin{equation}
\label{spin}
 m_i\to X_{ij}m_j\ ,
\end{equation}
where $X_{ij}\in SO(d-2)$ depends on the spacetime coordinate $x^\mu$.
Boosts are local rescaling of the null basis:
\begin{equation}
\label{boost}
 \ell\to \lambda\ell\ ,\quad n\to n/\lambda\ ,
\end{equation}
where $\lambda$ is any real scalar function. 
Null rotations about $\ell$ and $n$ are 
\begin{equation}
\label{nullrot1}
 \ell\to \ell,\quad
 n\to n+z_im_i-z^2\ell/2\ ,\quad
 m_i\to m_i-z_i\ell\ ,
\end{equation}
and 
\begin{equation}
\label{nullrot2}
 \ell\to \ell+z_i'm_i-z'{}^2n/2,\quad
 n\to n\ ,\quad
 m_i\to m_i-z_i'n\ ,
\end{equation}
where $z_i$ and $z_i'$ are real functions of $x^\mu$.

In the GHP formalism, we maintain 
the covariance with respective to spin and boost transformations.
An object $T_{i_1\cdots i_s}$ is a GHP scalar of spin $s$ and boost
weight $b$ if it transforms by the spins and boosts as
$T_{i_1\cdots i_s}\to X_{i_1j_1}\cdots X_{i_sj_s}T_{j_1\cdots j_s}$
and $T_{i_1\cdots i_s}\to \lambda^b T_{i_1\cdots i_s}$.
For example, the quantities $\rho_{ij},\tau_i,\kappa_i$ are GHP scalars
with $b=1,0,2$, respectively.
We also define priming operation: 
$T_{i_1\cdots i_s}\to T'_{i_1\cdots i_s}$,
where $T'_{i_1\cdots i_s}$ is the object obtained by exchanging $\ell$
and $n$ in the definition of $T_{i_1\cdots i_s}$.

We define GHP scalars obtained from Weyl tensor $C_{abcd}$ as
\begin{eqnarray}
&&\Omega_{ij}=C_{0i0j}\ ,\quad
\Omega'_{ij}=C_{1i1j}\ ,\\
&&\Psi_{ijk}=C_{0ijk}\ ,\quad
\Psi'_{ijk}=C_{1ijk}\ ,\quad
\Psi_{i}=C_{010i}\ ,\quad
\Psi'_{i}=C_{101i}\ ,\\
&&\Phi_{ij}=C_{0i1j}\ ,\quad
\Phi_{ijkl}=C_{ijkl}\ ,\quad
\Phi=C_{0101}\ ,\quad
\Phi^S_{ij}=\Phi_{(ij)}\ ,\quad
\Phi^A_{ij}=\Phi_{[ij]}\ ,
\end{eqnarray}
where $\Omega$, $\Psi$, $\Phi$, $\Psi'$ and $\Omega'$
have boost weights $b=2,1,0,-1,-2$, respectively.
The null vector $\ell$ is called multiple WAND (Weyl-aligned null
direction) iff all boost weight $+2$ and $+1$ components of the Weyl 
tensor vanish.
The spacetime admitting the multiple WAND is called algebraically
special spacetime. 
We can also obtain GHP scalars from 
Maxwell field strength $F_{ab}$ are
\begin{equation}
 \varphi_i=F_{0i}\ ,\quad
F=F_{01}\ ,\quad
F_{ij}=F_{ij}\ ,\quad
\varphi'_i=F_{1i}\ ,
\end{equation}
where $\varphi$, $F$ and $\varphi'$ 
have boost weights $b=1,0,-1$, respectively.

The partial derivatives of GHP scalars, such as
$\ell^\mu \partial_\mu T_{i_1\cdots i_s}$, 
$n^\mu \partial_\mu T_{i_1\cdots i_s}$ or 
$m_i^\mu \partial_\mu T_{i_1\cdots i_s}$, 
are not GHP scalars.
It is convenient to define derivative operators which are covariant
under spins and boosts as
\begin{eqnarray}
&&\tho T_{i_1\cdots i_s}=\ell^\mu \partial_\mu T_{i_1\cdots i_s}
-bL_{10}T_{i_1\cdots i_s}+\sum^s_{r=1}M^k_{i_r0}T_{i_1\cdots i_{r-1}ki_{r+1}\cdots i_s}\
,\\
&&\thop T_{i_1\cdots i_s}=n^\mu \partial_\mu T_{i_1\cdots i_s}
-bL_{11}T_{i_1\cdots i_s}+\sum^s_{r=1}M^k_{i_r1}T_{i_1\cdots i_{r-1}ki_{r+1}\cdots i_s}\
,\\
&&\es_i T_{j_1\cdots j_s}=m_{i}^\mu \partial_\mu T_{j_1\cdots j_s}
-bL_{1i}T_{j_1\cdots j_s}
+\sum^s_{r=1}M^k_{j_ri}T_{j_1\cdots j_{r-1}kj_{r+1}\cdots j_s}\
.
\end{eqnarray}
They are called GHP derivatives.
We can check that $\tho T_{i_1\cdots i_s}$, 
$\thop T_{i_1\cdots i_s}$ and $\es_i T_{j_1\cdots j_s}$
are all GHP scalars, with boost weight $(b+1,b-1,b)$ and spins
$(s,s,s+1)$.

The GHP scalars defined above are not independent because of 
Ricci equations, 
$[\nabla_\mu,\nabla_\nu]V_\rho=R_{\mu\nu\rho\sigma}V^\sigma$,
Bianchi equations, $\nabla_{[\lambda}C_{\mu\nu|\rho\sigma]}=0$,
and Maxwell equations, $dF=d\ast F=0$.
The relation for the GHP scalars in Einstein spacetimes 
$R_{\mu\nu}=\Lambda g_{\mu\nu}$ are summarized in
appendix.\ref{app:ghpeqns}.
Since these equations are invariant under the spins and boosts, 
they are written by GHP scalars and their GHP derivatives.

In the GHP formalism, the Klein-Gordon equation $(\nabla^2-\mu^2)\phi=0$
is written as
\begin{equation}
\label{full0}
 (2\thop\tho+\es_i\es_i+\rho'\tho-2\tau_i\es_i+\rho\thop-\mu^2)\phi=0\ .
\end{equation}
From appropriate linear combinations of equations in
appendix.\ref{app:ghpeqns}, 
we can obtain useful equations for studying electromagnetic and
gravitational perturbations~\cite{Durkee:2010qu}.
They are written as
\begin{equation}
\label{full}
\begin{split}
&(2\thop\tho+\es_j\es_j+\rho'\tho-4\tau_j\es_j+\Phi-\frac{2d-3}{d-1}\Lambda
)\varphi_i
+(-2\tau_i\es_j+2\tau_j\es_i+2\Phi^S_{ij}+4\Phi^A_{ij})\varphi_j\\
&=
[\kappa\thop 
+\rho\es 
+(\es \rho)
+(\thop\kappa)
+\rho\tau
+\kappa\rho'
+\Psi]F\\
&\qquad\qquad\qquad
+(\rho\thop
+\kappa\kappa'
+\rho\rho')\varphi
+(\kappa\es
+\rho^2
+\kappa\tau
+\Omega)\varphi'
\ ,
\end{split}
\end{equation}
and 
\begin{equation}
\label{graveq}
\begin{split}
&(2\thop\tho+\es_k\es_k+\rho'\tho-6\tau_k\es_k
+4\Phi-\frac{2d}{d-1}\Lambda)\Omega_{ij}\\
&\qquad\qquad\qquad\qquad
+4(\tau_k\es_{(i}-\tau_{(i}\es_k+\Phi^S_{(i|k}+4\Phi^A_{(i|k})\Omega_{i|j)}
+2\Phi_{ikjl}\Omega_{kl}+4\kappa_k\thop (\Psi_{(ij)k}+\Psi_{(i}\delta_{j)k})\\
&=[\rho\es+\tau\rho+\tau'\rho+\kappa\rho'+(\thop\kappa)+(\es\rho)+\Psi]\Psi
+\rho^2\Phi
+\kappa\rho\Psi'
+(\rho\thop\Omega+\rho\rho'+\kappa\kappa')\Omega
\ .
\end{split}
\end{equation}
The right hand sides of these equations are very long so we wrote them
schematically. 
In this paper, we do not need
the detailed expressions of the right hand sides.

\section{Extreme black holes in the GHP formalism}
\label{EXBHinGHP}

We consider general extreme black holes.
The metric of the extreme black holes
can be written as~\cite{Reall:2002bh}
\begin{equation}
 ds^2=L^2(x)[-r^2 F(r,x)dv^2 + 2dvdr]
+ \gamma_{\alpha \beta}(r,x)(dx^\alpha -r h^\alpha(r,x)dv)(dx^\beta -r
h^\beta(r,x)dv)\ ,
\label{metric1}
\end{equation}
where functions $(F,\gamma_{\alpha \beta},h^\alpha)$ and $L^2$ 
are smooth
function of $\{r,x^a\}$ and $\{x^a\}$, respectively.
The horizon of the spacetime is located at $r=0$.
In the metric, there is a residual coordinate transformation, 
$r\to \Gamma(x)r$. 
We choose the free function $\Gamma(x)$ so that $F(r=0,x)=1$ is
satisfied.
In this paper, we focus only on Einstein spacetimes satisfying
$R_{\mu\nu}=\Lambda g_{\mu\nu}$.

We assume that the background metric have $n$ rotational symmetry
generated by $\partial/\partial \phi^I$ $(I=1,2,\cdots,n)$. 
Then, the metric can be written as 
\begin{multline}
  ds^2=L^2(y)[-r^2 F(r,y)dv^2 + 2dvdr]
+ \gamma_{AB}(r,y)(dy^A -r h^A(r,y)dv)(dy^B -r h^B(r,y)dv)\\
+ 2\gamma_{AI}(r,y)(dy^A -r h^A(r,y)dv)(d\phi^I -r h^I(r,y)dv)\\
+ \gamma_{IJ}(r,y)(d\phi^I -r h^I(r,y)dv)(d\phi^J -r h^J(r,y)dv)
\ .
\label{metric2}
\end{multline}
We impose further assumption on metric functions as
\begin{equation}
 h^A(r,y)=\mathcal{O}(r)\ ,\qquad
 h^I(r,y)=k^I + \mathcal{O}(r)\ ,\qquad
 \gamma_{AI}(r,y)=\mathcal{O}(r)\ ,
\label{assm}
\end{equation}
where $k^I$ are constants. 
These assumptions are true 
for a large class of extreme black holes~\cite{Bardeen:1999px,Kunduri:2007vf,Figueras:2008qh,Kunduri:2008rs,Chow:2008dp}.
Under these assumptions, 
the near horizon geometry of the metric~(\ref{metric2}) 
takes ``standard'' form:
\begin{equation}
  ds^2=L^2(y)[-R^2dV^2 + 2dVdR]
+ \gamma_{AB}(y)dy^Ady^B\\
+ \gamma_{IJ}(y)(d\phi^I -r k^Idv)(d\phi^J -r k^Jdv)\ ,
\label{NHmetric}
\end{equation}
where we took the double scaling limit: $r=\epsilon R$, $v=V/\epsilon$ and $\epsilon\to 0$. 
The induced metric on the horizon is written as
\begin{equation}
\label{horind}
ds^2_H=\hat{g}_{\mu\nu}dx^\mu dx^\nu=\gamma_{AB}(y)dy^Ady^B+\gamma_{IJ}(y)d\phi^Id\phi^J\ .
\end{equation}

We take null basis $\{e_0,e_1,e_i\}=\{\ell,n,m_i\}$ 
in the general extreme black hole metric~(\ref{metric1}) as
\begin{equation}
\label{nullbasis}
\ell=\frac{2}{L}\partial_v +
 \frac{r^2F}{L}\partial_r+\frac{2rh^i}{L} \hat{e}_i\ ,
\quad
n=\frac{1}{2L}\partial_r\ ,
\quad
m_i =\hat{e}_i\ ,
\end{equation}
where $h^i\equiv h^\alpha \hat{e}^i_\alpha$ and 
$\hat{e}^i$ is an appropriate orthogonal basis for
$\gamma_{\alpha\beta}$. 
The null basis~(\ref{nullbasis}) is regular at the future horizon $r=0$.
Using the basis, we can obtain GHP variables. The full expression 
of the GHP variables are summarized in appendix.\ref{app:cal}. 
Here, we focus on $\rho_{ij}$ and $\kappa_i$
since they will be important later. They are given as
\begin{equation}
\kappa_i=\frac{2r^2F_{,i}}{L}\ ,\qquad
\rho_{ij}=
\frac{2r}{L}h^\alpha{}_{,(j}\hat{e}_{i) \alpha} 
+\frac{2r}{L}h^k \hat{e}^\alpha_{(j}(\hat{e}_{i)\alpha})_{,k}
+\frac{r^2F}{L}\hat{e}^\alpha_{(j}(\hat{e}_{i) \alpha})'
\ ,
\end{equation}
where ${}_{,i}\equiv \hat{e}^\mu_i\partial_\mu$.
We chose the residual gauge freedom so that $F(r=0,x)=1$ is
satisfied.
Thus, we obtain $F_{,i}=\mathcal{O}(r)$.
Therefore, we have $\kappa_i=\mathcal{O}(r^3)$.
From the assumption~(\ref{assm}), $h^\alpha|_{r=0}$ is constant and
we have $h^\alpha_{,j}=\mathcal{O}(r)$. 
Thus, the first term in $\rho_{ij}$ is $\mathcal{O}(r^2)$.
In the second term, there is a derivative operator, 
$h^k \partial_k=k^I \partial_{\phi^I} + \mathcal{O}(r)$.
Since the $\partial/\partial_{\phi^I}$ is a Killing vector, 
its operation to
background variables vanishes and the second term is also
$\mathcal{O}(r^2)$.
The last term is trivially $\mathcal{O}(r^2)$.
Therefore, we can conclude that $\rho_{ij}$ is second order in $r$.
By the similar way, we obtain the near horizon expression of the GHP
variables as
\begin{equation}
\label{NHGHP1a}
\begin{split}
&\rho_{ij}=\mathcal{O}(r^2)\ ,\qquad
\kappa_i=\mathcal{O}(r^3)\ ,\qquad
\tau_i=\frac{-(L^2)_{,i}+k_i}{2L^2}+ \mathcal{O}(r)
\ ,\\
&\rho'_{ij}=\mathcal{O}(1)\ ,\quad
\kappa'_i=0, \qquad
\tau'_i=\mathcal{O}(1)\ ,\\
&L_{10}=\frac{2r}{L}+\mathcal{O}(r^2)
\ ,\qquad
L_{11}=0\ ,\qquad
L_{1i}=\frac{k_i}{2L^2}+ \mathcal{O}(r)
\ ,\\
&M^i_{j0}=\mathcal{O}(r^2)
\ ,\qquad
M^i_{j1}=\mathcal{O}(1)\ ,\qquad
M^i_{jk}=\mathcal{O}(1)\ .
\end{split}
\end{equation}
Components of Weyl tensors with boost parameter $+2$ and $+1$ 
are given by Newman-Penrose equations~(\ref{NP1}) and (\ref{NP2}) as
\begin{equation}
\begin{split}
&\Omega_{ij}=-\tho\rho_{ij}+\es_j\kappa_i-\rho_{ik}\rho_{kj}-\kappa_i\tau'_j
-\tau_i\kappa_j=\mathcal{O}(r^3)\ ,\\
&\Psi_{ijk}=2(\tau_i\rho_{[jk]}+\kappa_i\rho'_{[jk]}-\es_{[j|}\rho_{i|k]})=\mathcal{O}(r^2)\ .
\end{split}
\label{PsiOme}
\end{equation}
Now, we consider 
Components of Weyl tensor with boost parameter $0$.
In~Ref.\cite{Durkee:2010xq}, it was shown that the induced Riemann tensor on
a spacelike surface which is orthogonal to null vectors $\ell$ and $n$ 
is written as
\begin{equation}
 R_{ijkl}^{(d-2)}=2\rho_{k[i|}\rho'_{l|j]}+2\rho'_{k[i|}\rho_{l|j]}+\Phi_{ijkl}
+\frac{2\Lambda}{d-1}\delta_{[i|k}\delta_{|j]l}\ .
\end{equation}
Thus, we have
\begin{equation}
\label{PhiNH}
\begin{split}
&\Phi_{ijkl}=\hat{R}_{ijkl}-\frac{2\Lambda}{d-1}\delta_{[i|k}\delta_{j]l}
+\mathcal{O}(r)\ ,\\
&\Phi_{ij}^S=-\frac{1}{2}(\hat{R}_{ij}-\frac{d-3}{d-1}\Lambda
 \delta_{ij})
+\mathcal{O}(r)\ ,\qquad
\Phi=-\frac{1}{2}(\hat{R}-\frac{(d-2)(d-3)}{d-1}\Lambda)
+\mathcal{O}(r)\ ,
\end{split}
\end{equation}
where $\hat{R}_{ijkl}$ is the induced Riemann tensor on the horizon~(\ref{horind}).
From the antisymmetric part of Eq.(\ref{NP4}), we obtain
\begin{equation}
\Phi_{ij}^A=\left[
-\frac{1}{4L^2}dk+\frac{1}{4L^2} dL^2\wedge k
\right]_{ij}+\mathcal{O}(r)\ .
\end{equation}
The GHP derivatives are 
\begin{equation}
\label{GHPdiff_ex}
\begin{split}
&\tho T_{i_1\cdots i_s}=\frac{2}{L}\{\partial_v  +r(k^I\partial_{\phi^I}-b)\} T_{i_1\cdots i_s} + \mathcal{O}(r^2)\ ,\\
&\thop T_{i_1\cdots i_s}=\frac{1}{2L}\partial_r T_{i_1\cdots i_s}+
\sum^s_{r=1}M^k_{i_r1}T_{i_1\cdots i_{r-1}ki_{r+1}\cdots i_s}\ ,\\
&\es_i T_{j_1\cdots j_s}=\left(\hat{\nabla}_i -
 \frac{bk_i}{2L^2}\right)T_{j_1\cdots j_s}
+ \mathcal{O}(r)\ ,
\end{split}
\end{equation}
where $\hat{\nabla}$ is a covariant derivative with respect to the
horizon induced metric~(\ref{horind}).

\section{Scalar field perturbations}
\label{scalar_pert}

\subsection{Conserved quantity on the horizon}
First, we consider the scalar field perturbation equation~(\ref{full0}).
The instability of the massless scalar field perturbation 
on any extreme black holes
has been already shown in~Ref.\cite{Lucietti:2012sf}. 
Here, we revisit the problem including the massive scalar field
using the GHP formalism.

Substituting near horizon expressions of 
GHP variables and derivatives~(\ref{NHGHP1a}-\ref{GHPdiff_ex}) into the
Klein-Gordon equation~(\ref{full0}),
we have the scalar field equation near the horizon as
\begin{equation}
\label{scalareq}
 \partial_v\big[2L(2\thop\phi+\rho'\phi)\big]
=\mathcal{A}_{0}\phi+\mathcal{O}(r)\ .
\end{equation}
where the operator $\mathcal{A}_0$ is defined as\footnote{
For axisymmetric perturbations, 
this operator $\mathcal{A}_0$ coincides with the operator 
$\mathcal{O}^{(0)}$ defined in the study of perturbation of near horizon
geometries~\cite{Durkee:2010ea}. 
We will also define operators  $\mathcal{A}_s$ for electromagnetic $(s=1)$
and gravitational $(s=2)$ perturbations by similar ways as the scalar
field, which 
relate to $\mathcal{O}^{(s)}$ defined in~Ref.\cite{Durkee:2010ea} 
as
$\mathcal{O}^{(1)}=\mathcal{A}_1$ and 
$\mathcal{O}^{(2)}=\mathcal{A}_2+2$ for axisymmetric perturbations.
}
\begin{equation}
\label{A0}
 \mathcal{A}_0\phi=-\hat{\nabla}_i(L^2\hat{\nabla}_i\phi)
+ik^Im_I\phi+\mu^2 L^2\phi\ ,
\end{equation}
where we decomposed $\{\phi^I\}$-dependence of $\phi$ by Fourier modes
$e^{im_I\phi^I}$, that is, $\partial_I \phi=m_I \phi$. We will do same decompositions for electromagnetic and
gravitational perturbations.
The derivation of the equation is given in
appendix.\ref{derNH}. 
Hereafter, we focus on axisymmetric perturbations $m_I=0$. 
For axisymmetric perturbations, the operator $\mathcal{A}_0$
is self-adjoint,
$(Y_1,\mathcal{A}_0Y_2)=(\mathcal{A}_0Y_1,Y_2)$, 
with respect to an inner product
\begin{equation}
 (Y_1,Y_2)=\int_H \  Y_1^\ast Y_2 \ ,
\end{equation}
where $\int_H=\int d^{d-2}x \sqrt{\hat{g}}$ and $\hat{g}$ is the
determinant of the horizon
induced metric defined in Eq.(\ref{horind}).

We assume that operator $\mathcal{A}_0$ has a zero eigenvalue.
This is always true for massless case $\mu=0$ since we have
$\mathcal{A}_0Y=0$ when $Y$ is a constant. 
For massive scalar field, 
$\mathcal{A}_0$ can also have zero eigenvalues 
depending on the value of $\mu^2$ and background geometries.
We will discuss the existence of the zero eigenvalues in
section.\ref{eigenv}.
We denote the eigenfunction for the zero eigenvalue as $Y$.
Operating $(Y,*)$ to both side of Eq.(\ref{scalareq}), 
we obtain
\begin{equation}
\label{I0}
 \frac{dI_0}{dv}=0\ ,\qquad
I_0=\int_H Y^\ast \big[2L(2\thop\phi+\rho'\phi)\big]\ .
\end{equation}
where we used $(Y,\mathcal{A}_0\phi)=(\mathcal{A}_0Y,\phi)=0$.
Hence, $I_0$ is a conserved quantity along the horizon.
Note that the GHP derivative $\thop$ contains only the radial
derivative $\partial_r$. (See Eq.(\ref{GHPdiff_ex}).)
Thus, the integrand in the $I_0$ 
is written by the linear combination of $\partial_r\phi$ and $\phi$.
Therefore, we can conclude that if $I_0\neq 0$ at an initial surface, 
$\partial_r \phi$ and $\phi$ do not both decay along
the future horizon as $v\to \infty$.

\subsection{Instability against scalar field perturbations}

We assume that 
$\phi$ and its angular derivatives $\partial /\partial y^A$ 
decay along the horizon. 
For extreme Reissner-Nordstr\"{o}m and Kerr black holes,
it was shown that $\phi$ decays along the horizon.
So, this assumption seems likely also for any other extreme black holes.
Then, at the late time, the conserved quantity $I_0$ approaches
\begin{equation}
 I_0\simeq 2\int_H Y^\ast \partial_r \phi\ ,\quad(v\to \infty)\ .
\end{equation}
Now, we differentiate Eq.(\ref{full0}) by $r$.
Near the horizon, the equation can be written as
\begin{equation}
\label{diffr0}
  \partial_v\big[2L\partial_r(2\thop\phi+\rho' \phi)\big]
=(\mathcal{A}_0-2)\partial_r\phi+\mathcal{D}_0\phi+\mathcal{O}(r)\ ,
\end{equation}
where, in the linear operator $\mathcal{D}_0$, there is 
no radial derivative $\partial_r$.
Thus, we have $\mathcal{D}_0\phi\to 0$ as $v\to \infty$.
The derivation of above equation is in appendix.\ref{derNH}.
Operating $(Y,*)$ to both side of Eq.(\ref{diffr0}), we obtain
\begin{equation}
 \frac{dJ_0}{dv}\simeq -I_0\ ,\quad(v\to \infty)\ ,\qquad
J_0=\int_H Y^\ast \big[2L\partial_r(2\thop\phi+\rho'\phi)\big]
\ 
\end{equation}
Therefore, the quantity $J_0$ blows up linearly in time $v$ as
\begin{equation}
 J_0\simeq -I_0v\ ,\quad(v\to \infty)\ .
\end{equation}
The integrand in the $J_0$ 
is written by the linear combination of $\partial_r^2\phi$,
$\partial_r\phi$ and $\phi$. Since we assumed that $\phi$ decays at the
horizon, either $\partial_r^2\phi$ or $\partial_r\phi$ 
blows up at the horizon.
This implies the instability of the extreme black holes against scalar
field perturbations.

\section{Electromagnetic perturbations}
\label{em_pert}

\subsection{Decoupled equation on the horizon}
\label{DecMax}

Secondly, we consider electromagnetic perturbations.
We consider the Maxwell field as test field and, thus, 
it vanishes in the background.
It follows that GHP scalars obtained from Maxwell field strength 
are all invariant under the infinitesimal coordinate
transformations and basis transformations~(\ref{spin})-(\ref{nullrot2}).
The number of physical degrees of freedom of the Maxwell field in
$d$-dimensions is $d-2$. The number of components of $\varphi_i$ is also
$d-2$. Thus, we can expect that $\varphi_i$ has all physical degrees
of freedom of electromagnetic perturbations and 
it would be nice if we can obtain  decouple equations for $\varphi_i$.
The right hand side of Eq.(\ref{full}) contains 
coupling terms between $\varphi_i$ and
other components of the perturbation. So, $\varphi_i$ does not decouple in
general. 
We can see that, however, all the terms in right hand side of 
Eq.(\ref{full}) are multiplied by  
$\rho$, $\kappa$, $\Omega$ or $\Psi$. 
From Eqs.(\ref{NHGHP1a}) and (\ref{PsiOme}), they are at most $\mathcal{O}(r^2)$.
Thus, the right hand side is $\mathcal{O}(r^2)$.
Therefore, in Eq.(\ref{full}) and its radial derivative,
the right hand side is zero at
the horizon.

\subsection{Conserved quantity on the horizon}

We consider left hand side of Eq.(\ref{full}) neglecting the right hand
side.
Using near horizon expressions of 
GHP variables and derivatives~(\ref{NHGHP1a}-\ref{GHPdiff_ex}), 
we obtain
\begin{equation}
 \partial_v [2L(2\thop\varphi_i+\rho'\varphi_i)]
=\mathcal{A}_1\varphi_i
+\mathcal{O}(r)\ .
\label{delv}
\end{equation}
where we define the operator $\mathcal{A}_1$ as
\begin{equation}
\label{A1}
\begin{split}
 \mathcal{A}_1\varphi_i
&= 
- \frac{1}{L^2} \hat{\nabla}_j ( L^4 \hat{\nabla}_j \varphi_i)
+\left(
2+3ik^Im_I-\frac{5}{4L^2}k_I k^I
\right)\varphi_i
+ L^2 \left(\hat{R}_{ij} + \frac{1}{2} \hat{R} g_{ij}\right)\varphi_j\\
& \qquad\qquad
+ \left(
-\frac{1}{2} (dk)_{ij} + 2(k-d(L^2))_{[i}
\hat{\nabla}_{j]} 
-\frac{1}{L^2}(dL^2)_{[i} k_{j]}
\right)\varphi_j\ .
\end{split}
\end{equation}
The derivation of the equation is written in appendix.\ref{derNH}.
Hereafter, we focus on the axisymmetric perturbations $m_I=0$.
We define an inner product as
\begin{equation}
 (Y^1,Y^2)=\int_H L^2 (Y_i^{1})^\ast Y^2_i \ .
\end{equation} 
For axisymmetric perturbations,
we have $(Y^1,\mathcal{A}_1Y^2)=(\mathcal{A}_1Y^1,Y^2)$, that
is, the operator $\mathcal{A}_1$ is self-adjoint.
We assume that
the operator $\mathcal{A}_1$ has zero eigenvalue.
Although the existence of the zero eigenvalue is not obvious,
we will see that many extreme black holes satisfy this assumption in
section.\ref{eigenv}.
We denote the eigenfunction by $Y_i$.
Operating $(Y,\ast)$ to Eq.(\ref{delv}),
we have
\begin{equation}
\label{I1}
 \frac{dI_1}{dv}=0\ ,\quad
I_1=\int_H Y_i^\ast [2L^3(2\thop\varphi_i+\rho'\varphi_i)]
\ ,
\end{equation}
where we used $(Y,\mathcal{A}_1\varphi)=(\mathcal{A}_1Y,\varphi)=0$.
Therefore, $I_1$ is a conserved quantity along the horizon.
Thus, If $I_1\neq 0$ at an initial surface, 
$\partial_r \varphi_i$ and $\varphi_i$ do not both decay along
the future horizon.

We 
assume that $\varphi_i$ and its tangential derivatives along the
horizon decay as $v\to\infty$. 
Now, we differentiate Eq.(\ref{full}) by $r$.
Then, the right hand side becomes $\mathcal{O}(r)$ 
and still vanishes on the horizon. 
Thus, we have
\begin{equation}
\label{diffr1}
  \partial_v\big[2L\partial_r(2\thop\varphi_i+\rho'\varphi_i)\big]
=\mathcal{A}_1\partial_r\varphi_i+\mathcal{D}_1\varphi_i+\mathcal{O}(r)\ ,
\end{equation}
where, in the linear operator $\mathcal{D}_1$, there is 
no radial derivative $\partial_r$.
Hence, we have $\mathcal{D}_1\varphi_i\to 0$ ($v\to \infty$).
The derivation of the equation is in appendix.\ref{derNH}.
Operating $(Y,\ast)$ to above equation and taking limit of $v\to\infty$, 
we obtain
\begin{equation}
 \frac{dJ_1}{dv}\simeq0\ ,\quad(v\to\infty)\ ,\qquad
J_1=\int_H Y_i^\ast [2L^3\partial_r(2\thop\varphi_i+\rho'\varphi_i)]
\end{equation}
Therefore, the quantity $dJ_1/dv$ 
tends to be zero at the late time even if we consider initial
data with $I_1\neq 0$.
So, we cannot show the instability of extreme black holes 
against electromagnetic perturbations by the same way as scalar
fields.
We may be able to find instability in 
higher order derivatives $\partial_r^n \varphi_i$ $(n>2)$.
However, 
since the coupling terms are $\mathcal{O}(r^2)$ in the equation,
we cannot neglect these terms in the limit of $r\to 0$ when we
consider the higher order radial derivatives of Eq.(\ref{full}).
It seems to be difficult problem to show the instability taking into
account the coupling terms.

\section{Gravitational perturbations}
\label{grav_pert}

\subsection{Gauge invariant variables on the horizon}
\label{gauge_inv}

Finally, we study gravitational perturbations.
We consider perturbation of GHP variables as 
$\Omega_{ij}\to \Omega_{ij}+\tilde{\Omega}_{ij}$,
$\Psi_{ijk}\to \Psi_{ijk}+\tilde{\Psi}_{ijk}$, etc. 
Here, variables with tildes represent first order perturbations.
Variables without tildes are background variables.
The number of physical degrees of freedom of the gravitational
perturbations is $d(d-3)/2$. 
On the other hand, 
the number of components of $\tilde{\Omega}_{ij}$ is also
$d(d-3)/2$. 
Thus, we can expect that $\tilde{\Omega}_{ij}$ 
has all physical degrees
of freedom of the gravitational
perturbations. 
The perturbation variable $\tilde{\Omega}_{ij}$ 
is transformed by gauge transformations 
as follows:\\
Coordinate transformations ($x^\mu\to x^\mu+\xi^\mu(x)$):
\begin{equation}
\label{CoordOm}
 \tilde{\Omega}_{ij}\to \tilde{\Omega}_{ij}+\xi^\mu \partial_\mu
  \Omega_{ij}\ .
\end{equation}
Spins ($t_{ij}\in so(d-2)$):
\begin{equation}
\label{SpinOm}
 \tilde{\Omega}_{ij}\to \tilde{\Omega}_{ij}+2it_{(i|k}\Omega_{k|j)}\ .
\end{equation}
Boosts:
\begin{equation}
\label{BoostOm}
 \tilde{\Omega}_{ij}\to \tilde{\Omega}_{ij}+2\alpha\Omega_{ij}\ .
\end{equation}
Null rotations:
\begin{equation}
\label{NrotOm}
 \tilde{\Omega}_{ij}\to \tilde{\Omega}_{ij}
-2z_k(\Psi_{(i}\delta_{j)k}+\Psi_{(ij)k})\ .
\end{equation}
Here, $\xi^\mu$, $t_{ij}$, $\alpha$ and $z_k$ are infinitesimal
functions depending on spacetime coordinates.
In Eq.(\ref{PsiOme}), 
we obtained $\Omega_{ij}=\mathcal{O}(r^3)$ and 
$\Psi_{ijk}=\mathcal{O}(r^2)$.
Thus, under spin and boost transformations,
we have $\tilde{\Omega}_{ij}\to \tilde{\Omega}_{ij}+\mathcal{O}(r^3)$.
On the other hand, under coordinate transformations and null rotations, 
we have $\tilde{\Omega}_{ij}\to \tilde{\Omega}_{ij}+\mathcal{O}(r^2)$.
Therefore, 
$\tilde{\Omega}_{ij}|_{r=0}$ and $\partial_r \tilde{\Omega}_{ij}|_{r=0}$ are
gauge invariant, but $\partial_r^2 \tilde{\Omega}_{ij}|_{r=0}$ is not
gauge invariant in general.
Thus, even if we could 
show that $\partial_r^2 \tilde{\Omega}_{ij}|_{r=0}$ blows
up along the horizon by the similar way as scalar field perturbations,
we cannot determine if the instability is physical one 
or just a gauge mode.
We will avoid this problem by assuming that the
background geometry is algebraically special in section.\ref{NrottoW}.

\subsection{Decoupled equations on the horizon}
\label{DecGrav}

We consider the first order perturbation of Eq.(\ref{graveq}).
The right hand side contains coupling term between 
$\tilde{\Omega}_{ij}$ and
other perturbation variables.
We can see that all terms in the right hand side are 
$\mathcal{O}(r^2)$.
For example, we have
$(\tau'\rho\Psi)\,\tilde{}=\tilde{\tau}'\rho\Psi+\tau'\tilde{\rho}\Psi+\tau'\rho\tilde{\Psi}=\mathcal{O}(r^2)$ 
since $\rho$ and $\Psi$ are the second order in $r$.
Thus, in Eq.(\ref{graveq}) and its radial derivative, 
the right hand side is zero at the horizon.
In the left hand side, there is another coupling term, 
$4\kappa_k\thop (\Psi_{(ij)k}+\Psi_{(i}\delta_{j)k})$
which can be expanded as
\begin{equation}
[4\kappa_k\thop (\Psi_{(ij)k}+\Psi_{(i}\delta_{j)k})]\,\tilde{}
=4\tilde{\kappa}_k\thop
  (\Psi_{(ij)k}
+\Psi_{(i}\delta_{j)k})
+\mathcal{O}(r^3)\ .
\end{equation}
This coupling term is $\mathcal{O}(r)$. 
(Recall that the GHP derivative $\thop$ contains 
the radial derivative $\partial_r$. Hence, we have $\thop\Psi_{ijk}=\mathcal{O}(r)$.)
Such a coupling term is harmless when we construct a conserved quantity
at the horizon and we can show the non-decay of the gravitational
perturbations by the same way as scalar and electromagnetic perturbations. 
However, when we prove that the perturbations blow up along the horizon,
the coupling term is problematic since we differentiate 
Eq.(\ref{graveq}) by $r$. 
In section.\ref{NrottoW},
we will see that this problem can also be avoided 
by assuming that the background geometry is algebraically special.

\subsection{Conserved quantity on the horizon}
\label{grav_conv}

The coupling terms in Eq.(\ref{graveq}) is
$\mathcal{O}(r)$ and negligible near the horizon.
Thus, near horizon, the equation becomes
\begin{equation}
 \partial_v \big[2L(2\thop\tilde{\Omega}_{ij}+\rho'\tilde{\Omega}_{ij})\big]
=\mathcal{A}_2\tilde{\Omega}_{ij}+\mathcal{O}(r)\ .
\label{delv_grav}
\end{equation}
where the operator $\mathcal{A}_2$ is defined as
\begin{equation}
\label{A2}
\begin{split}
&\mathcal{A}_2\tilde{\Omega}_{ij}=
-\frac{1}{L^4}\hat{\nabla}_k(L^6\hat{\nabla}_k\tilde{\Omega}_{ij})
+\left(4+3ik^Im_I-\frac{4k^Ik_I}{L^2}-2(d-4)\Lambda L^2\right)\tilde{\Omega}_{ij}\\
&+2L^2\left(\hat{R}_{(i|k}+\hat{R}\delta_{(i|k}\right)\tilde{\Omega}_{k|j)}
-2L^2\hat{R}_{ikjl}\tilde{\Omega}_{kl}\\
&+\left[
-(dk)_{(i|k}-\frac{2}{L^2}(d(L^2)\wedge k)_{(i|k}
+2(k-d(L^2))_{(i|}\hat{\nabla}_{k}
-2(k-d(L^2))_{k}\hat{\nabla}_{(i|}\right]\tilde{\Omega}_{k|j)}\ .
\end{split}
\end{equation}
The derivation of the equation is written in appendix.\ref{derNH}.
For axisymmetric perturbations $m_I=0$, the operator $\mathcal{A}_2$
is self-adjoint with respect to an inner product
\begin{equation}
 (Y^1,Y^2)=\int_H L^4 Y^1{}^\ast_{ij} Y^2_{ij}\ .
\end{equation}
Hereafter, we focus on axisymmetric perturbations.
We assume that operator $\mathcal{A}_2$ has a zero eigenvalue.
In section.\ref{eigenv}, we will see that many extreme black holes
satisfy this assumption.
We denote the eigenfunction for the zero eigenvalue as $Y_{ij}$.
Operating $(Y,*)$ to both side of Eq.(\ref{delv_grav}), 
we obtain
\begin{equation}
\label{I2}
 \frac{dI_2}{dv}=0\ ,\qquad
I_2=\int_H Y^\ast_{ij} \big[2L^5(2\thop\tilde{\Omega}_{ij}+\rho'\tilde{\Omega}_{ij})\big]\ .
\end{equation}
where we used $(Y,\mathcal{A}_2\tilde{\Omega})=(\mathcal{A}_2Y,\tilde{\Omega})=0$.
Therefore, $I_2$ is a conserved quantity along the horizon.
Thus, If $I_2\neq 0$ at an initial surface, 
$\partial_r \tilde{\Omega}_{ij}$ and $\tilde{\Omega}_{ij}$ do not both decay along
the future horizon as $v\to \infty$.
Recall that both of $\partial_r \tilde{\Omega}_{ij}$ and $\tilde{\Omega}_{ij}$ are gauge
invariant at the horizon.

\subsection{Null rotation to a multiple WAND}
\label{NrottoW}

As explained in 
section.\ref{gauge_inv} and \ref{DecGrav},
we require conditions $\Psi_{ijk}=\mathcal{O}(r^3)$ 
and $\Omega_{ij}=\mathcal{O}(r^4)$ to show the instability of the
gravitational perturbations. 
To satisfy  these conditions, 
we assume that the background geometry is
algebraically special. 
Then, there is a null rotation~(\ref{nullrot2}) which transforms 
the null vector $\ell$ to a multiple WAND, that is, 
$\Psi_{ijk}=\Omega_{ij}=0$.
In the near horizon geometry~(\ref{NHmetric}), 
we have already known the multiple WAND: 
$\ell^{NH}=L^{-1}(2\partial_V+r^2\partial_R+2rk^I\partial_\phi^I)$.
We can expect that,
in the near horizon limit:  
$r=\epsilon R$, $v=V/\epsilon$ and $\epsilon\to 0$,
the multiple WAND in the full geometry coincides with 
the $\ell^{NH}$ modulo boost transformations.\footnote{
We checked that this is correct for Kerr-NUT-AdS spacetimes~\cite{Chen:2006xh}.
That is, the multiple WAND in Kerr-NUT-AdS spacetimes found in
Ref.\cite{Hamamoto:2006zf} approaches $\ell^{NH}$ in the near horizon limit. 
}
The null vector $\ell$ defined in Eq.(\ref{nullbasis}) satisfies this
condition by itself. ($\ell\to \epsilon \ell^{NH}$ in the near horizon
limit.) Thus, 
$z_i'$ in the null rotation~(\ref{nullrot2}) 
should be $\mathcal{O}(r^2)$.
It follow that the near horizon expressions of GHP variables~(\ref{NHGHP1a}) are correct even after
the this null rotation.

\subsection{Instability against gravitational perturbations}

We assume that $\tilde{\Omega}_{ij}$ 
and its tangential derivatives along the
horizon decay along the horizon. 
Then, at late time, the conserved quantity $I_2$ becomes 
\begin{equation}
 I_2\simeq 2\int_H L^4 Y^\ast_{ij} \partial_r \tilde{\Omega}_{ij}\ ,\quad(v\to \infty)\ .
\end{equation}
Now, we differentiate Eq.(\ref{graveq}) by $r$.
Since we assumed that the background geometry is algebraically special,
we can neglect the coupling term in the equation.
Near the horizon, the equation can be written as
\begin{equation}
\label{diffr2}
  \partial_v\big[2L\partial_r(2\thop\tilde{\Omega}_{ij}+\rho'\tilde{\Omega}_{ij})\big]
=(\mathcal{A}_2+2)\partial_r\tilde{\Omega}_{ij}+\mathcal{D}_2\tilde{\Omega}_{ij}+\mathcal{O}(r)\ ,
\end{equation}
where, in the linear operator $\mathcal{D}_2$, 
there is no radial derivative.
Thus, we have $\mathcal{D}_2\tilde{\Omega}_{ij}\to 0$ ($v\to \infty$).
The derivation of above equation is in appendix.\ref{derNH}.
Operating $(Y,*)$ to both side of Eq.(\ref{diffr0}), we obtain
\begin{equation}
 \frac{dJ_2}{dv}\simeq I_2\ ,\quad(v\to \infty)\ ,\qquad
J_2=\int_H Y^\ast_{ij} \big[2L^5\partial_r(2\thop\tilde{\Omega}_{ij}+\rho'\tilde{\Omega}_{ij})\big]
\ 
\end{equation}
Therefore, the quantity $J_2$ blow up linearly in time $v$ as
\begin{equation}
 J_2\simeq I_2v\ ,\quad(v\to \infty)\ .
\end{equation}
Thus, either $\partial_r^2\tilde{\Omega}_{ij}$ or
$\partial_r\tilde{\Omega}_{ij}$ blows up along the horizon.
This implies the instability of the extreme black holes against
gravitational perturbations.

\section{Unstable extreme black holes}
\label{UnsEX}
\subsection{Summary of our statement}
\label{SmSt}

Our statements obtained in this paper are as follows:
If the operator $\mathcal{A}_s$ has a zero eigenvalue for
an axisymmetric perturbation and 
the horizon conserved quantity $I_s$ is non-zero, 
$\partial_r \psi_s$ and $\psi_s$ do not both decay along
the future horizon as $v\to \infty$,
where $\psi_0=\phi$, $\psi_1=\varphi_i$ and
$\psi_2=\tilde{\Omega}_{ij}$. 
The explicit expressions of $\mathcal{A}_s$ are given in Eqs.(\ref{A0}),
(\ref{A1}) and (\ref{A2}).
The horizon conserved
quantities~(\ref{I0}), (\ref{I1}) and (\ref{I2}) are written as
\begin{equation}
 I_s=\int_H Y^\ast \cdot
  \big[2L^{2s+1}(2\thop\psi_s+\rho'\psi_s)\big]\ ,
\end{equation}
where $Y$ is the eigenfunction satisfying $\mathcal{A}_sY=0$.
In the proof of this statement, we used assumptions~(\ref{assm}).

Hereafter, we assume that 
$\psi_s$ and its tangential derivatives along the
horizon decay as $v\to\infty$. 
(Then, $\partial_r \psi_s$
cannot decay.)
For scalar field perturbations, 
we can show that 
either $\partial_r^2\phi$ or $\partial_r\phi$
blows up along the horizon.
For gravitational perturbations, 
when the background geometry is algebraically special,
either $\partial_r^2\tilde{\Omega}_{ij}$ or
$\partial_r\tilde{\Omega}_{ij}$ 
also blows up along the horizon.
For electromagnetic perturbations, 
we could not find the instability in $\partial_r^2\varphi_{i}$ or $\partial_r\varphi_{i}$. 
(There may be instability in the higher order derivative by $r$.)

\subsection{Eigenvalues of $\mathcal{A}_s$}
\label{eigenv}

The existence of a zero eigenvalue for the horizon operator 
$\mathcal{A}_s$ is crucial for the proof of the instability.
Surprisingly, 
in study of perturbations of near horizon geometries,
the eigenvalues of $\mathcal{A}_s$ 
have been calculated
for some extreme black holes:
4-dimensional extreme Kerr black holes~\cite{Dias:2009ex,Amsel:2009ev},
all 5-dimensional black holes with two rotational 
symmetries for $\Lambda=0$~\cite{Murata:2011my}
and Myers-Perry(-AdS) black holes with equal angular
momenta~\cite{Durkee:2010ea,Tanahashi:2012si}. 
In this subsection, 
we investigate the existence of zero eigenvalues of $\mathcal{A}_s$ 
using their results.

For massless scalar field, 
the existence of the zero eigenvalue is trivial since 
we have $\mathcal{A}_0Y=0$ when $Y$ is a constant.
For massive case, there is no zero eigenvalue in general.
In some cases, however,
it has a zero eigenvalue depending on mass and background geometry.
For example, in odd-dimensional Myers-Perry-AdS spacetimes with equal
angular momenta, 
it was shown that the eigenvalue $\lambda_0$ is given by~\cite{Durkee:2010ea}
\begin{equation}
 \frac{\lambda_0}{L^2}=\frac{4\kappa(\kappa+N)}{r_+^2}+\mu^2\ ,\quad
  (\kappa=0,1,2,\cdots)\ .
\end{equation}
where $r_+$ is the horizon radius and $L^2$ is a constant.
The integer $N$ relates to the spacetime dimension as $d=2N+3$.
From this expression, when the scalar field mass
is given by $\mu^2=-4\kappa(\kappa+N)/r_+^2$, 
the operator $\mathcal{A}_0$ has a
zero mode. 
(If the horizon radius $r_+$ is sufficiently large, 
the $\mu^2$ does not violate the Breitenl\"{o}hner-Freedman bound.)

For electromagnetic and gravitational perturbations ($s=1,2$), 
the existence of a zero eigenvalue is not obvious.
In 4-dimensional extreme Kerr geometry,
the operator $\mathcal{A}_s$ ($s=1,2$) 
does not have zero eigenvalue. 
However, if we consider perturbation equations 
for $\Omega'_{ij}$ instead of $\Omega_{ij}$,
we can show the instability~\cite{Lucietti:2012sf}. 
(In the 4-dimensional Kerr geometry, $\Omega'_{ij}$ satisfies a decoupled
equation although it does not decouple in Myers-Perry spacetimes even
if we consider the near horizon limit.)
For all 5-dimensional black holes with two rotational 
symmetries, eigenvalues $\lambda_s$ are written as\footnote{
Note that we need to shift the eigenvalues when we use 
the results in Refs.\cite{Durkee:2010ea,Murata:2011my,Tanahashi:2012si} 
for gravitational perturbations since 
the operator $\mathcal{A}_2$ relates to $\mathcal{O}^{(2)}$ defined by
them as $\mathcal{O}^{(2)}=\mathcal{A}_2+2$ 
for axisymmetric perturbations.
}
\begin{equation}
\begin{split}
&\lambda_1=\ell(\ell+1)\ ,\  (\ell+1)(\ell+2)\ ,\  (\ell+1)(\ell+2)\ ,\\
&\lambda_2=(\ell-1)(\ell+2)\ ,\ \ell(\ell+3), \ \ell(\ell+3)\ ,\ (\ell+1)(\ell+3)\ ,\ (\ell+1)(\ell+3)\ .
\end{split}
\end{equation}
where $\ell=0,1,2,\cdots$. We can find that 
$\lambda_1$ and $\lambda_2$ can be zero for $\ell=0$ and
$\ell=0,1$, respectively. 
Thus, in these spacetimes,
the gravitational and electromagnetic 
perturbations do not decay in general.
In particular, for 5-dimensional Myers-Perry black holes,
we can show that gravitational perturbations 
(either $\partial_r^2\tilde{\Omega}_{ij}|_{r=0}$ or
$\partial_r\tilde{\Omega}_{ij}|_{r=0}$)
blow up along the horizon 
since the spacetimes are known to be algebraically
special~\cite{Pravda:2007ty}. 

The horizon induced metrics of Myers-Perry black holes with equal angular
momenta can be viewed as Hopf fibration over $CP^N$ where $N$ is the
integer part of $(d-3)/2$.
Thus, the eigenfunction of $\mathcal{A}_s$ can be
decomposed into tensor, vector and scalar harmonics on the base space
$CP^N$. 
All eigenvalues for axisymmetric modes are given in
Refs.\cite{Durkee:2010ea,Tanahashi:2012si}. 
In scalar modes, we can always find zero eigenvalues. 
(Tensor and vector modes can also have zero eigenvalues depending on the
spacetime dimension $d$.)
Thus, Myers-Perry black holes with equal angular
momenta in all dimensions are unstable against gravitational
perturbations. (Electromagnetic perturbations do not decay at least.)
Therefore, as far as we calculated, 
all vacuum extreme higher dimensional black holes with vanishing
cosmological constant 
have zero eigenvalues
in the horizon operator $\mathcal{A}_s$.
It would be nice if we can show the existence of the zero eigenvalues
for general black holes.

\section{Discussions}
\label{Conc}

We studied 
perturbations in general extreme black hole spacetimes in all dimensions.
We found a sufficient condition for instability which is
summarized in section.\ref{SmSt}.
Using the condition, 
we showed that 5-dimensional extreme Myers-Perry black holes are unstable against
gravitational perturbations.
For $d\geq 6$, we also found gravitational instability in 
extreme Myers-Perry black holes
when they have equal angular momenta.

In the study of perturbations of near horizon
geometries~\cite{Durkee:2010ea,Murata:2011my,Tanahashi:2012si},
they considered dimensional reduction of perturbation equations and
obtained effective equations of motion in AdS$_2$.
Then, they used a criterion $m^2<-1/4$
to determine the instability of
near horizon geometry, where 
$m$ is the effective mass in the AdS$_2$, 
since this implies violation of 
the Breitenl\"{o}hner-Freedman (BF) bound.
The effective mass relates to eigenvalue of
$\mathcal{A}_s$ as 
$m^2=\lambda_0$, $\lambda_1$, $\lambda_2+2$ for
scalar, electromagnetic and gravitational perturbations, respectively.
We can see that the condition of the instability obtained in this paper
($\lambda_s=0$) differs from theirs. 
This discrepancy comes from difference of type of instabilities.
The violation of BF bound ($m^2<-1/4$) 
is considered as a condition for an exponential
grow of the perturbations. This was explicitly shown for scalar field
perturbations~\cite{Durkee:2010ea}.
On the other hand, our condition $\lambda_s=0$ gives a 
power law grow of the perturbations, 
which is more modest than the exponential one.
For ($d\geq 6$)-dimensional extreme Myers-Perry black holes 
with equal angular momenta, 
it was shown that the BF bound in the near horizon geometries
is violated for gravitational 
perturbations~\cite{Durkee:2010ea,Tanahashi:2012si}.
In fact, in the case odd-dimensions, 
such an instability has been found in the full 
geometry near the extremality~\cite{Dias:2010eu}.
So, the power law instability found in this paper may not be important
for these spacetimes.
However, for 5-dimensional extreme Myers-Perry black holes 
with equal angular momenta, 
there is no violation of the BF bound in the near horizon geometries.
In addition to that, from the study of perturbations of full geometries, 
strong evidence of stability for non-extreme black holes has been found in~Ref.\cite{Murata:2008yx}.
Thus, the power law instability found in this paper can be important for
this spacetime.

In this paper, we considered only vacuum black holes. Thus, our
instability condition
does not apply to gravitational and electromagnetic perturbations of
Reissner-Nordstr\"{o}m (RN) or Kerr-Newman black
holes. 
Further work needs to be done to study their stability in extreme limit.
They are solutions of $N=2$ supergravity and extreme RN black holes are supersymmetric.
For RN black holes, it is known that the perturbation equations are
decoupled. Using the decoupled equations, we may be able to find 
conserved quantities on the horizons and show the instability~\cite{LMRT}.
For Kerr-Newman black holes, the decoupling of the perturbation
equations has not been succeeded. Hence, even for the non-extreme case, 
their stability has not been studied.
As we did in this paper, however, 
if the coupling terms in the perturbation equations 
are sufficiently small near the horizons,
we can study the instability.
It would be interesting to estimate the order of the coupling terms and 
study the instability of 
extreme Kerr-Newman black hole.
It would make  good progress in understanding of stability of Kerr-Newman
black holes.

One of the most interesting problems on the instability is its final 
state.  
We need to solve the time evolution of the 
instability taking into account the backreaction to specify the final state. 
For scalar field perturbations of RN black holes, we can find an 
instability for spherically symmetric modes~\cite{Aretakis:2011hc}. 
Thus, the evolution equations of the instability are given by 
$(1+1)$-dimensional partial differential equations even if we consider 
the backreaction. Solving the PDEs and finding the final state  
would be another direction of the future research.

\section*{Acknowledgments}
We would like to thank 
to H.~Reall for collaboration at an earlier stage and careful reading of
the paper.
KM is supported by JSPS Grant-in-Aid for
Scientific Research No.24$\cdot$2337.
This work is partially supported by European Research Council grant 
no. ERC-2011-StG 279363-HiDGR.
\appendix

\section{GHP variables for extreme black holes}
\label{app:cal}

We take the null basis as in Eq.(\ref{nullbasis}).
The dual one-forms for these vectors are written as
\begin{equation}
 e^0=e_1=\frac{L}{2}dv\ ,\quad
 e^1=e_0=-r^2LFdv+2Ldr\ ,\quad
 e^i=e_i=e^i-rh^idv\ .
\end{equation}
Using the Cartan equations $de_a+\omega_{cab}e^c\wedge e^b=0$,
we can calculate the spin connections $\omega_{abc}$.
From the definition of $L_{ab}$ in Eq.(\ref{LNMdef}),
we have $L_{ab}=-\omega_{b0a}$.
Thus, GHP variables 
for the background spacetime are given as
\begin{equation}
\begin{split}
&\rho_{ij}=
\frac{2r}{L}h^\alpha{}_{,(j}\hat{e}_{i) \alpha} 
+\frac{2r}{L}h^k \hat{e}^\alpha_{(j}(\hat{e}_{i)\alpha})_{,k}
+\frac{r^2F}{L}\hat{e}^\alpha_{(j}(\hat{e}_{i) \alpha})'
\ ,\\
&\kappa_i=\frac{2r^2F_{,i}}{L}
\ ,
\qquad
\tau_i=-\frac{L_{,i}}{L}+\frac{(rh^\alpha)'\hat{e}^i_\alpha}{2L^2}
\ ,\\
&\rho'_{ij}=\frac{1}{2L}(\hat{e}_{\alpha
 (i})'\hat{e}^\alpha_{j)}\ ,
\qquad
\kappa'_i=0, 
\qquad
\tau'_i=-\frac{L_{,i}}{L}-\frac{(rh^\alpha)'\hat{e}^i_\alpha}{2L^2}
\ ,
\end{split}
\end{equation}
and
\begin{equation}
\begin{split}
&L_{10}=\frac{(r^2F)'}{L}+\frac{2rh^iL_{,i}}{L^2}
\ ,
\qquad
L_{11}=0\ ,
\qquad
L_{1i}=
\frac{(rh^\alpha)'\hat{e}^i_\alpha}{2L^2}
\ ,\\
&M^i_{j0}=
\frac{2r}{L}h^\alpha{}_{,[j}\hat{e}_{i] \alpha} 
+\frac{2r}{L}h^k \hat{e}^\alpha_{[j}(\hat{e}_{i]\alpha})_{,k}
+\frac{r^2F}{L}\hat{e}^\alpha_{[j}(\hat{e}_{i] \alpha})'
\ ,\\
&M^i_{j1}=
\frac{1}{2L}(\hat{e}_{\alpha
 [i})'\hat{e}^\alpha_{j]}\ ,
\qquad
M^i_{jk}=-\hat{\omega}_{kij}\ ,
\end{split}
\end{equation}
where $\hat{\omega}_{kij}$ is defined by 
$d\hat{e}_i+\hat{\omega}_{kij}\hat{e}^k\wedge \hat{e}^j=0$.

\section{Derivation of near horizon equations}
\label{derNH}

Here, we derive near horizon equations~(\ref{scalareq}), 
(\ref{diffr0}), (\ref{delv}), (\ref{diffr1}), (\ref{delv_grav}) 
and (\ref{diffr2}).
The equations for scalar, electromagnetic and gravitational
perturbations are written in a unified form as
\begin{equation}
\label{unify}
 (2\thop\tho+\rho'\tho)\psi_s
+\mathcal{B}_s\psi_s=0\ ,
\end{equation}
where $\psi_0=\phi$, $\psi_1=\varphi_i$ and
$\psi_2=\tilde{\Omega}_{ij}$. 
The angular operators $\mathcal{B}_s$ are defined by
\begin{align}
&\mathcal{B}_0\psi_0=(\es_i\es_i-2\tau_i\es_i+\rho\thop-\mu^2)\phi\ ,\\
&\mathcal{B}_1\psi_1=(\es_j\es_j-4\tau_j\es_j+\Phi-\frac{2d-3}{d-1}\Lambda
)\varphi_i
+(-2\tau_i\es_j+2\tau_j\es_i+2\Phi^S_{ij}+4\Phi^A_{ij})\varphi_j\ ,\\
&\mathcal{B}_2\psi_2=(\es_k\es_k-6\tau_k\es_k
+4\Phi-\frac{2d}{d-1}\Lambda)\tilde{\Omega}_{ij}\notag\\
&\qquad\qquad\qquad\qquad
+4(\tau_k\es_{(i}-\tau_{(i}\es_k+\Phi^S_{(i|k}+4\Phi^A_{(i|k})\tilde{\Omega}_{i|j)}
+2\Phi_{ikjl}\tilde{\Omega}_{kl}\ .
\end{align}
In $\mathcal{B}_s$, there is no the radial derivative
$\partial_r$.
Up to second order in $r$, 
the GHP derivative $\tho$ can be written as
\begin{equation}
 \tho\psi_s=\frac{2}{L}[\partial_v+r(ikm-b)]\psi_s
+\frac{r^2}{L}\partial_r\psi_s+r^2\mathcal{C}\psi_s
+\mathcal{O}(r^3)\ ,
\end{equation}
where $\mathcal{C}$ is a operator in which the radial derivative
$\partial_r$ is not contained.
From Eq.(\ref{GHPdiff_ex}), we obtain $[\thop,r]=1/(2L)$.
Thus, we have
\begin{equation}
\begin{split}
 (2\thop\tho+\rho'\tho)\psi_s
&=\partial_v\left[\frac{2}{L}(2\thop\psi_s+\rho'\psi_s)\right]
+\frac{2}{L^2}(ikm-b)\psi_s\\
&\qquad+\frac{2r}{L}\left\{
2(ikm-b)\thop+\frac{1}{L}\partial_r+\mathcal{C}'\right\}\psi_s
+\mathcal{O}(r^2)\ .
\end{split}
\end{equation}
where $\mathcal{C}'=\mathcal{C}+\rho'(ikm-b)$.
We expand the operator $\mathcal{B}_s$ as
$\mathcal{B}_s=\mathcal{B}_s^H+r\mathcal{B}_s^1+\mathcal{O}(r^2)$.
Then, Eq.(\ref{unify}) is written as
\begin{equation}
\label{psieq}
\begin{split}
&\partial_v\left[\frac{2}{L}(2\thop\psi_s+\rho'\psi_s)\right]
+\frac{2}{L^2}(ikm-b)\psi_s
+\mathcal{B}_s^H \psi_s\\
&\qquad\qquad
+\frac{2r}{L}\left\{
2(ikm-b)\thop+\frac{1}{L}\partial_r+\mathcal{C}''\right\}\psi_s
=\mathcal{O}(r^2)\ ,
\end{split}
\end{equation}
where $\mathcal{C}''=\mathcal{C}'+L\mathcal{B}_s^1/2$.
Thus, from Eq.(\ref{psieq}), we obtain
\begin{equation}
\label{unify0}
 \partial_v\left[2L(2\thop\psi_s+\rho'\psi_s)\right]
=\mathcal{A}_s\psi_s
+\mathcal{O}(r)\ ,
\end{equation}
where 
\begin{equation}
\mathcal{A}_s=-2(ikm-b)-L^2\mathcal{B}_s^H\ . 
\end{equation}
Eq.(\ref{unify0}) expresses Eq.(\ref{scalareq}), (\ref{delv}) and (\ref{delv_grav}).
Differentiating Eq.(\ref{psieq}) by $r$, we have
\begin{equation}
 \partial_v\left[2L\partial_r(2\thop\psi_s+\rho'\psi_s)\right]
=[\mathcal{A}_s+2(b-1)-2ikm]\partial_r\psi_s
+\mathcal{C}'''\psi_s
+\mathcal{O}(r)\ ,
\end{equation}
where $\mathcal{C}'''=\mathcal{C}''+2(ikm-b)(\thop-\partial_r/(2L))$.
(Note that there is no radial derivative in the operator $\thop-\partial_r/(2L)$.) 
Setting $m_I=0$ in above equation, we obtain 
Eqs.(\ref{diffr0}), (\ref{diffr1}) and (\ref{diffr2}). 
The explicit expressions of $\mathcal{A}_s$ 
can be obtained using 
near horizon expressions of GHP variables and
derivatives~(\ref{NHGHP1a}), (\ref{PhiNH}) and (\ref{GHPdiff_ex}).

\section{Useful GHP equations}\label{app:ghpeqns}
We summarize the useful GHP equations for Einstein spacetime satisfying
$R_{\mu\nu}=\Lambda g_{\mu}$.
These equations are firstly derived in Ref.\cite{Durkee:2010xq}.

\subsection{Newman-Penrose equations}
From the Ricci equations, 
$[\nabla_\mu,\nabla_\nu]V_\rho=R_{\mu\nu\rho\sigma}V^\sigma$, we obtain
following equations.
\begin{align}
\tho \rho_{ij} - \es_j \kappa_i &= - \rho_{ik} \rho_{kj} -\kappa_i
\tau'_j - \tau_i \kappa_j - \Omega_{ij},\tag{NP1}\label{NP1}\\
 \tho \tau_i - \thop \kappa_i &= \rho_{ij}(-\tau_j + \tau'_j) -
 \Psi_i,\tag{NP2}\label{NP2}\\
 2\es_{[j|} \rho_{i|k]} &= 2\tau_i \rho_{[jk]} + 2\kappa_i \rho'_{[jk]}
 - \Psi_{ijk}
 ,\tag{NP3}\label{NP3}\\
 \tho' \rho_{ij} - \es_j \tau_i &= - \tau_i \tau_j - \kappa_i \kappa'_j
 - \rho_{ik}\rho'_{kj}-\Phi_{ij}
 - \frac{\Lambda}{d-1}\delta_{ij}.\tag{NP4}\label{NP4}
\end{align}
Another four equations can be obtained by taking the prime $'$ of these
four.

\subsection{Bianchi equations}\label{sec:bianchi}
From Bianchi equations, $\nabla_{[\lambda}C_{\mu\nu|\rho\sigma]}=0$,
we obtain following equations.\\
{\noindent\bf Boost weight +2:}
\begin{align}
  \tho \Psi_{ijk} - 2 \es_{[j}\Omega_{k]i} 
&= (2\Phi_{i[j|} \delta_{k]l} - 2\delta_{il}
\Phi^A_{jk}-\Phi_{iljk})\kappa_l \notag \\
& -2 (\Psi_{[j|} \delta_{il} + \Psi_i\delta_{[j|l} + \Psi_{i[j|l} 
+ \Psi_{[j|il}) \rho_{l|k]} + 2 \Omega_{i[j} \tau'_{k]},\tag{B1}\label{B1}
\end{align}
{\bf Boost weight +1:}
\begin{align}
  - \tho \Phi_{ij} - \es_{j}\Psi_i + \thop \Omega_{ij} 
=& - (\Psi'_j \delta_{ik} - \Psi'_{jik}) \kappa_k + (\Phi_{ik} + 2\Phi^A_{ik} + \Phi \delta_{ik}) \rho_{kj} 
                      \notag\\
&  + (\Psi_{ijk}-\Psi_i\delta_{jk}) \tau'_k - 2(\Psi_{(i}\delta_{j)k} + \Psi_{(ij)k}) \tau_k 
                     - \Omega_{ik} \rho'_{kj}, \tag{B2}\label{B2}\\
  -\tho \Phi_{ijkl} + 2 \es_{[k}\Psi_{l]ij}
                 =& - 2 \Psi'_{[i|kl} \kappa_{|j]} - 2
		 \Psi'_{[k|ij}\kappa_{|l]} \notag\\
&  + 4\Phi^A_{ij} \rho_{[kl]} -2\Phi_{[k|i}\rho_{j|l]} 
                     + 2\Phi_{[k|j}\rho_{i|l]} + 2
		     \Phi_{ij[k|m}\rho_{m|l]} \notag\\
                 &  -2\Psi_{[i|kl}\tau'_{|j]} - 2\Psi_{[k|ij} \tau'_{|l]}
                     - 2\Omega_{i[k|} \rho'_{j|l]} + 2\Omega_{j[k} \rho'_{i|l]},
                     \tag{B3}\label{B3}\\
  -\es_{[j|} \Psi_{i|kl]}
                 =& 2\Phi^A_{[jk|} \rho_{i|l]} - 2\Phi_{i[j} \rho_{kl]} 
                     + \Phi_{im[jk|} \rho_{m|l]} - 2\Omega_{i[j} \rho'_{kl]},\tag{B4}\label{B4}
\end{align}
{\bf Boost weight 0:}
\begin{align}
  \thop \Psi_{ijk} -2 \es_{[j|}\Phi_{i|k]} 
                 =& 2(\Psi'_{[j|} \delta_{il} - \Psi'_{[j|il}) \rho_{l|k]}
                     + (2 \Phi_{i[j}\delta_{k]l} - 2\delta_{il}\Phi^A_{jk} - \Phi_{iljk}) \tau_l \nonumber\\
                 &  + 2 (\Psi_i \delta_{[j|l} -  \Psi_{i[j|l})\rho'_{l|k]} + 2\Omega_{i[j}\kappa'_{k]},
                     \tag{B5}\label{B5}\\
  -2\es_{[i} \Phi^A_{jk]} 
                 =& 2\Psi'_{[i} \rho_{jk]} + \Psi'_{l[ij|} \rho_{l|k]} 
                     - 2\Psi_{[i} \rho'_{jk]} - \Psi_{l[ij|} \rho'_{l|k]},\tag{B6}\label{B6}\\
  -\es_{[k|} \Phi_{ij|lm]} 
                 =& - \Psi'_{i[kl|} \rho_{j|m]} + \Psi'_{j[kl|} \rho_{i|m]} 
                     - 2\Psi'_{[k|ij} \rho_{|lm]}\nonumber\\
                  & - \Psi_{i[kl|} \rho'_{j|m]} + \Psi_{j[kl|} \rho'_{i|m]} 
                     - 2\Psi_{[k|ij} \rho'_{|lm]}.\tag{B7}\label{B7}
\end{align}
Another five equations are obtained 
by applying the prime operator to above equations.

\subsection{Maxwell equations}
From Maxwell equations, $dF=d\ast F=0$, 
we obtain following equations.
\begin{align}
  \es_i \varphi_i + \tho F\tag{M1}\label{max1}
         =& \tau'_i \varphi_i + \rho_{ij} F_{ij} - \rho F - \kappa_i \varphi'_i \\
  2 \es_{[i} \varphi_{j]} - \tho F_{ij} \tag{M2}\label{max2}
         =& 2\tau'_{[i} \varphi_{j]} + 2F \rho_{[ij]}
             + 2F_{[i|k} \rho_{k|j]} + 2\kappa_{[i}\varphi'_{j]}\\
  2\thop \varphi_i + \es_j F_{ji} - \es_i F
         =& (2\rho'_{[ij]}-\rho' \delta_{ij}) \varphi_j
                      - 2F_{ij} \tau_j - 2 F\tau_i
                      + (2\rho_{(ij)}-\rho \delta_{ij}) \varphi'_j \tag{M3}\label{max3}\\
  \es_{[i} F_{jk]} \tag{M4}\label{max4}
         =& \varphi_{[i} \rho'_{jk]} + \varphi'_{[i} \rho_{jk]}
\end{align}
A further three equations can be obtained by priming above equations.

\subsection{Commutators of derivatives}
The commutation relations for GHP derivatives are given by
\begin{align}
\left[\tho, \thop \right]T_{i_1...i_s} 
=& (-\tau_j + \tau'_j) \es_jT_{i_1...i_s} 
+b\left(-\tau_j\tau'_j + \kappa_j\kappa'_j + \Phi \right)T_{i_1...i_s} \notag\\
&+ \sum_{r=1}^s \left(\kappa_{i_r} \kappa'_{j} - \kappa'_{i_r} \kappa_{j} 
  + \tau'_{i_r} \tau_{j} - \tau_{i_r} \tau'_{j} + 2\Phi^A_{i_r j}\right) T_{i_1...j...i_s}, \tag{C1}\label{C1}\\
\left[\tho, \es_i\right] T_{k_1...k_s}
=& -(\kappa_i \thop + \tau'_i\tho +\rho_{ji}\es_j)T_{k_1...k_s}
+ b\left(-\tau'_j\rho_{ji} + \kappa_j\rho'_{ji} + \Psi_i \right)T_{k_1...k_s} \notag\\
+&  \sum_{r=1}^s \left( \kappa_{k_r}\rho'_{li} - \rho_{k_r i}\tau'_l
+ \tau'_{k_r} \rho_{li} - \rho'_{k_r i} \kappa_l - \Psi_{ilk_r}\right) T_{k_1...l...k_s}, \tag{C2}\label{C2}\\
\left[\es_i,\es_j\right] T_{k_1...k_s}
=& \left(2\rho_{[ij]} \tho' + 2\rho'_{[ij]} \tho \right) T_{k_1...k_s}
+ b \left(2\rho_{l[i|} \rho'_{l|j]} + 2\Phi^A_{ij}\right) T_{k_1...k_s}\notag\\
& + \sum_{r=1}^s \left(2\rho_{k_r [i|} \rho'_{l|j]} + 2\rho'_{k_r [i|} \rho_{l|j]} 
+ \Phi_{ijk_r l} + \frac{2\Lambda}{d-1} \delta_{[i|k_r}\delta_{|j]l} 
                           \right) T_{k_1...l...k_s}. \tag{C3}\label{C3}
\end{align}
The result for $[\thop, \eth_i]$ can be obtained from 
the prime operation of $[\tho, \eth_i]$.

\end{document}